\documentclass[12pt, twocolumn]{iopart}
\usepackage{amsfonts}
\usepackage{amssymb}
\usepackage[dvips]{graphicx}
\usepackage[normal]{caption}
\begin{document}
\paper{Falling spring and falling catenary as cases of force propagation}

\author{Hyunbyuk Kim$^{1,\ast}$ and Kyoungdae Kim$^{1,2,\dagger}$}

\address{$^1$ Department of Physics and Earth Science, Korea Science Academy of KAIST, 899, Danggam 3-dong, Busanjin-gu, Busan, Korea}
\address{$^2$ Korea Advanced Institute of Science and Technology (KAIST), 373-1 Guseong-dong, Yuseong-gu, Daejeon 305-701, Korea}
\ead{$^\ast$time@kaist.ac.kr and $^\dagger$kkim@kaist.ac.kr}

\begin{abstract}
At every points of a static equilibrium system, the net force is zero. If one of the composite forces of this system is disappeared, it is no more in equilibrium and this effect of absence spreads through the system with a finite velocity. So it takes finite time before this absence is manifest at the other points. A frequently asked question may describe this situation well. ``If the Sun disappears abruptly, does this affect us instantly?" In this question centripetal force due to the gravity of the Sun does not disappear simultaneously all over the place, so the Earth sustains its elliptical motion for a while. As this example shows, it is a well known fact that the force does spread with a finite velocity. But for the specific problems it is sometimes not easy to notice this property of the force. To help this, we consider the motion of a falling spring and a falling catenary. We study the conceptual aspects of these motions and apply it to the experimental data. A minimal theoretical background is also treated. For the future, we anticipate that this work would be useful for enhancing the concept of wave motion.
\end{abstract}
\maketitle

\section{Introduction}
The force does not act instantaneously over a distance. Especially the contact force on the material shows this fact well. For example when a ball gets hit, the opposite side of the hit point feels the impulse a little bit later than the moment of the hit \cite{ref:baseball}. Of course, this can be observed only through a high speed camera and we usually ignore this fact at undergraduate courses. Moreover this finiteness of force transferring speeds, which is usually mentioned once in dealing with the field force, barely used in problem solving thereafter. As a field force transferring phenomenon, we treat the retarded potential in the electrodynamics \cite{ref:retarded1}. Likewise, gravity, as it is a field force, transferring involves elaborated calculations and this cannot be easily done \cite{ref:retarded2}. That is why we do not attack the ``abrupt absence of the Sun" problem in the general physics course. But for the contact force, as we will see, this is not necessarily so. So here is one subtle point in the physics education of this general property of the force. That is students learn this property but they are not asked to use it. The object of this paper is to clear this subtle point by studying the concrete examples.

Main reason of not using this property at undergraduate courses is that most problems can be dealt with a particle model. In other words, when we draw the motion diagram we replace the object with a single point and ignore the internal motion. To use the particle model, however, the size of the object must be much less than the length scales of the problem \cite{ref:particle}. Except when dealing with the elasticity, this approximation is so frequently used such that the students nearly automatically see the problems with this particle prejudices. They tend to see the bulk as a particle. So even they know the fact that the force requires time to be transferred over material or over vacuum, they implicitly neglect this.

The particle approach is inevitable for the introductory courses.
But the motion of the bulk object cannot be fully analyzed with this particle approximation only. If we overly insist particle approach, sometimes we miss the physical features of the problem. Especially the features related with the finiteness of the force transferring speeds. As clear examples of force transferring phenomenon which cannot be anticipated with the particle model, we consider the falling spring and the falling catenary.

Particularly, physics of the falling spring has attracted many authors \cite{ref:spring1, ref:spring2, ref:spring3}. There are two fascinating features of this motion. One is the levitation of the bottom point and the other is constant speed (not constant acceleration) falling of the top point. When this problem is given to the students, their educated guess is not that quite correct. And when the correct answer is given, they usually show hard to believe reactions. To help this situation, we revisit the falling spring and discuss the physical implications of the problem.

This paper is organized as follows. In theory section we treat the motion of the falling spring and the falling catenary by using the analytic and numerical method respectively. We will see the similarity and dissimilarity between these two motions. But still the core physical arguments are intact. Then the experimental result of the falling spring is given at the demonstration section. We will see how well the data are compatible with the physical arguments. Finally, the discussion about the property of the force and enlarged view for undergraduate courses will close this paper.

\section{Theory}
\subsection{The falling spring}

\begin{figure}
\centering
\includegraphics[width=0.5\textwidth]{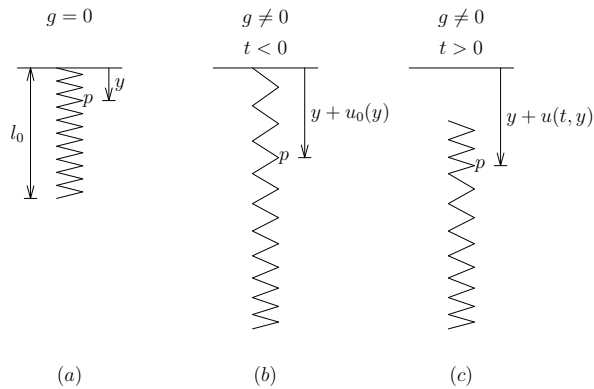}
\caption{A freely falling spring.
(a) Without gravity, the spring has unstretched length $l_0$. $-l_0 \leq y \leq 0$.
(b) Before $t=0$, the spring is stretched due to the gravity and then it is in the static equilibrium. $u_0(y) \leq 0$.
(c) At $t=0$, the spring begins to fall.}
\label{fig:springsetup}
\end{figure}

Calkin had given the analytical solution of the falling spring \cite{ref:spring2}. For the purpose of illustration and future use, we give here the sketch of that analysis. There are two different approximation schemes. One is a loosely wound spring and the other is a tightly wound spring. Both schemes predict same qualitative behaviour. As you will notice at the demonstration section our experimental set up is more appropriate for the tightly wound spring case. But for the simplicity and clarity we follow here loosely wound scheme. Please consult Calkin's paper and references therein for a full treatment.

A spring of mass $m$, spring constant $k$ and unstretched length $l_0$ is suspended from its top end and hanging vertically in static equilibrium under the influence of the Earth's gravitational field $-g$. At time $t=0$ the top end is released and the spring falls. The problem is to find its subsequent motion. See \fref{fig:springsetup} for the notations and conventions \cite{ref:spring2, ref:spring3, ref:springsetup}.

\begin{figure}
\centering
\includegraphics[width=0.8\textwidth]{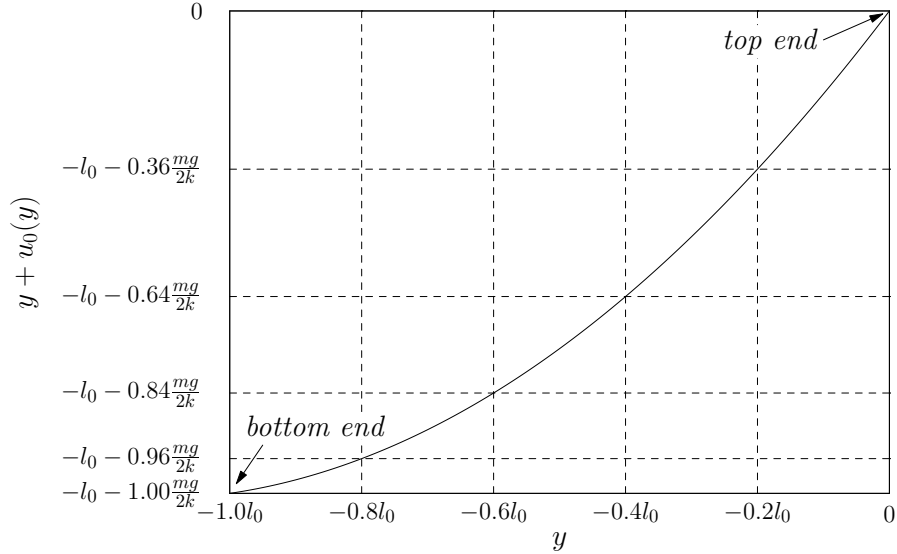}
\caption{Initially the spring is in equilibrium. Its stretched length due to the gravity is $l_0+\frac{mg}{2k}$. Note the mass density increases at the lower part of the spring.}
\label{fig:springinit}
\end{figure}

Before the spring falls, its top point is fixed
\begin{equation}
u_0(y=0)=0,
\end{equation}
and its bottom point has no tension
\begin{equation}
E \left( \frac{\partial u_0(y)}{\partial y} \right)_{y=-l_0}=0,
\end{equation}
where $E$ is Young's modulus, because there is no hanging mass under the spring. At each point of the spring the composite forces are partial weight of the spring under that point (i.e., $w = g \mu (l_0+y)$ where $\mu$ is a line density of mass.) and the restoring tension due to the elasticity. They balance each other when the spring is in static equilibrium. So the Newton's equation is
\begin{equation}\label{eq:eqofmotionspring}
\mu \frac{\partial^2 u(t,y)}{{\partial t}^2} = E \frac{\partial^2 u(t,y)}{{\partial y}^2} - \mu g.
\end{equation}
Before the falling, there is no motion so the equation becomes
\begin{equation}
{\left( l \sqrt{k/m}\right)}^2  \frac{\partial^2 u(t,y)}{{\partial y}^2} - g = 0 \quad (\textrm{for} \,\,\, t < 0),
\end{equation}
where we have used $E=k l$ and $\mu = m/l$. With the help of the initial and boundary conditions we can get the position of each point on the spring
\begin{equation}
y + u_0(y) = y + \frac{mg}{2k} \left( 2\frac{y}{l_0} + \frac{y^2}{{l_0}^2} \right) \quad (\textrm{for} \,\,\, t < 0).
\end{equation}
So the initially stretched length is given by
\begin{equation}\label{eq:springl}
l = l_0 + \frac{mg}{2k}.
\end{equation}
See \fref{fig:springinit}.

At the moment of starting to fall $t=0$, each point on the spring maintains its initial position 
\begin{equation}
u(t=0, y)=u_0(y) = \frac{mg}{2k} \left( 2\frac{y}{l_0} + \frac{y^2}{{l_0}^2} \right).
\end{equation}
And their initial speeds are zero
\begin{equation}
\left( \frac{\partial u(t,y)}{\partial t} \right)_{t=0} = 0.
\end{equation}
With these, we can solve the Newton's equation for the subsequent falling motions. Finally the solution of the motion becomes
\begin{eqnarray}\label{eq:springsol}
\fl
y+u(t,y) = \left\{
\begin{array}{lrcl}
y + \frac{mg}{2k} \left( 2\frac{y}{l_0} + \frac{y^2}{{l_0}^2} \right) &
0 \leq & \sqrt{\frac{k}{m}} t & \leq -\frac{y}{l_0} \\
y + \frac{mg}{2k} \left( 2\frac{y}{l_0} + \frac{y^2}{{l_0}^2} \right) + \frac{mg}{k} \left( \sqrt{\frac{k}{m}} t + \frac{y}{l_0} \right) & 
\quad -\frac{y}{l_0} \leq & \sqrt{\frac{k}{m}} t & \leq \left( 2 +\frac{y}{l_0} \right) \\
\cdots & & \cdots & .
\end{array}
\right.
\end{eqnarray}

\begin{figure}
\centering
\includegraphics[bb=20 210 500 540, width=0.8\textwidth]{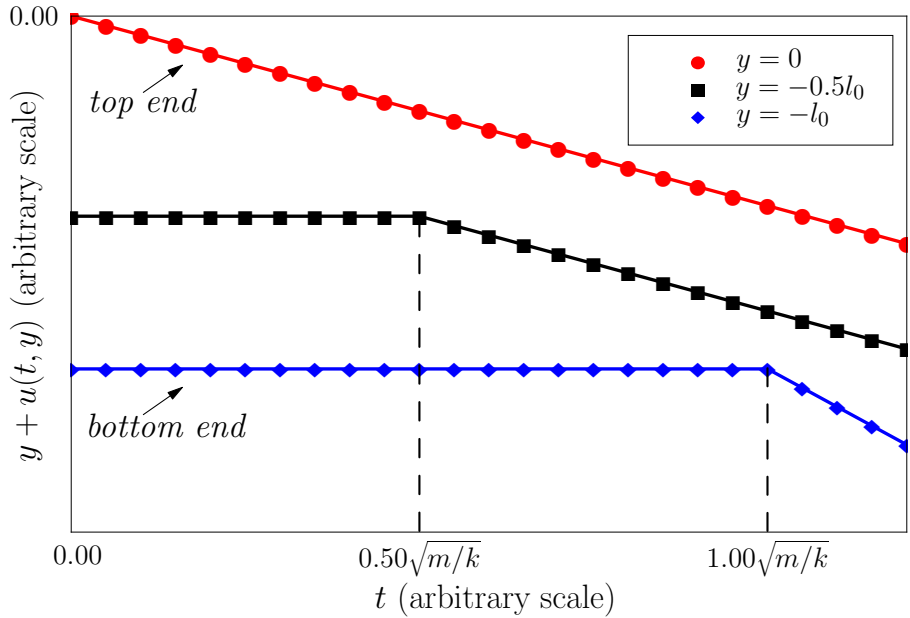}
\caption{Analytical solution of the falling spring. The graph is drawn with mock data. Until the wave arrives, each part of the spring maintains its current motion. Then it falls with a constant speed.}
\label{fig:springfall}
\end{figure}

When this situation is given, students come up with various speculations. To name a few, ``All points fall with constant acceleration g.", ``The top point is accelerating downward larger than g and the bottom point is accelerating downward smaller than g.", ``The bottom point accelerates upward first than downward." and ``The center of mass falls with a constant acceleration g.". From these trials, we can see the habit of seeing the problem with the particle model. But to fully taste the problem we must recognize the bulk as a continuum with finite size and the force takes time to transfer from one point to another.

\Fref{fig:springinit} describes the initial position of each point on the spring under the gravity. Note that the mass density increases at the lower part of the spring. \Fref{fig:springfall}, which is a plot of \eref{eq:springsol}, describes the subsequent motion of the each point of the spring. Notice two distinctive features of the motion. First, the bottom point levitates for a while. Second, the top point falls with constant velocity (not with the constant acceleration) with respect to the observer on the ground. As the top point approaches the bottom point, the bottom point starts to fall. After the bottom point starts to fall, the spring's motion becomes superposition of incident wave and reflected wave. This requires more advanced mathematics and beyond the scope of this paper \cite{ref:springsol}. We concentrate here only on the motion just before the falling of the bottom point.

This motion is rather counter intuitive to the first encounter. The key point to the understanding is that the application or the negation of the force propagates with a finite velocity through the system. In this falling spring case, due to the configuration of the set up, the propagation manifests itself in a form of longitudinal wave. And loosely speaking, the signal speed becomes the force propagation speed. (Since \eref{eq:eqofmotionspring} is a inhomogeneous wave equation, it is difficult to define the wave speed. Instead we use the term signal speed as a quotient of a length scale and the characteristic time scale of the system.) Till this information arrives, the point maintains its current motion. That's why all the points levitate during the finite time intervals respectively. And the bottom point shows this more dramatically. This signal speed is a kind of limiting speed of the propagation. For the longitudinal motion, the material which composes the solid bulk cannot move faster than the force transferring signal which propagates through the bulk, otherwise it would mean the breakthrough of the material without disturbing the system. This is a kind of acausality. As there should be no violation of causality, the material's moving speed should be capped by the signal speed. And that's why the top point falls with constant velocity unless the signal speed is increasing. (More accurate term will be a finite velocity. But this detail is not bothering us here.) A slightly modified question may help to understand this motion. What if the spring was stretched and then release not vertically but horizontally on the table?

\subsection{The falling catenary}

\begin{figure}
\centering
\includegraphics[width=0.5\textwidth]{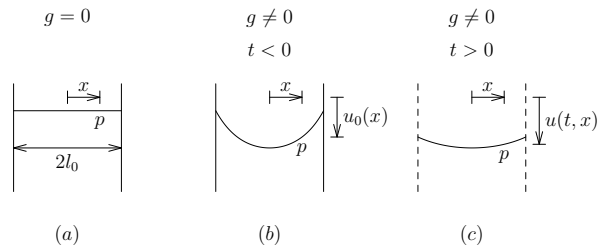}
\caption{A freely falling catenary.
(a) Without gravity, the rope has unstretched length $2l_0$. $-l_0 \leq x \leq l_0$.
(b) Before $t=0$, the rope is stretched due to the gravity and then it is in the static equilibrium. $u_0(x) \leq 0$.
(c) At $t=0$, the rope begins to fall.}
\label{fig:catenarysetup}
\end{figure}

Have equipped with the solution of the falling spring, we ask a similar question. Consider the hanging cord with its two end points are fixed at the same height. Its mass is $m$ and the width between the fixed points is $2 l_0$. It is initially at rest and at $t=0$ its two end points are set to be free under the influence of the Earth's gravitational field $-g$. What is its subsequent motion? Now you can guess that the middle point of the cord will levitate for a while and then fall. Let's check whether this is true or not.

The initial shape of the cord is called ``catenary" and given by
\begin{equation}
u_0(x) = l_0 \frac{T_H}{mg} \left[ \cosh\left({\frac{mg}{T_H} \frac{x}{l_0}}\right) - \cosh\left(\frac{mg}{T_H}\right) \right],
\end{equation}
where $T_H$ is a horizontal component of the tension \cite{ref:catenarysetup}.
See \fref{fig:catenarysetup} for the notations and conventions. At $t=0$ it starts to fall and the Newton's equation becomes wave equation with the additional gravitational acceleration
\begin{equation}
\frac{\partial^2 u(t,x)}{{\partial t}^2} = c^2 \frac{\partial^2 u(t,x)}{{\partial x}^2} - g,
\end{equation}
where $c=\sqrt{T/\mu}$ ($T$ is a tension. $\mu$ is a line density of the mass.).
The initial conditions are
\begin{equation}
u(t=0,x)=u_0(x),
\end{equation}
\begin{equation}
\left( \frac{\partial u(t,x)}{\partial t} \right)_{t=0} = 0.
\end{equation}
The boundary conditions are
\begin{equation}
u(t,x=\pm l_0) = -\frac{g}{2}t^2,
\end{equation}
because there were no artificially added horizontal forces.

\begin{figure}
\centering
\includegraphics[bb=20 210 530 520, width=0.8\textwidth]{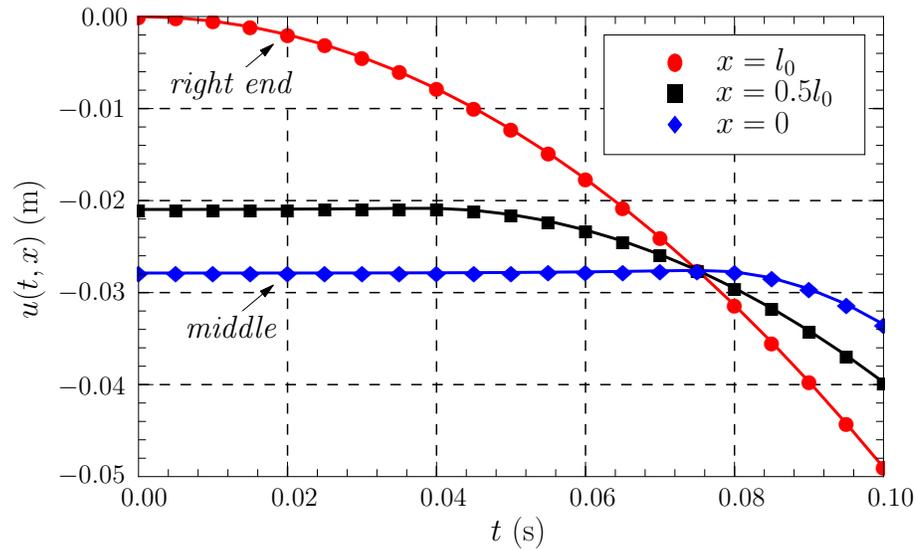}
\caption{Numerical results of the falling catenary. Until the signal arrives, each part of the catenary maintains its current motion. Then it falls with a constant acceleration. Used parameters are $g=9.8$m/$\textrm{s}^2$, $m=0.03\textrm{kg}$, $T_H=0.8\textrm{N}$, $l_0=0.15\textrm{m}$ and $c \simeq \sqrt{T_H l_0 / m} = 2.0\textrm{m/s}$.}
\label{fig:catenaryfall1}
\end{figure}

\begin{figure}
\centering
\includegraphics[bb=20 140 520 450, width=0.8\textwidth]{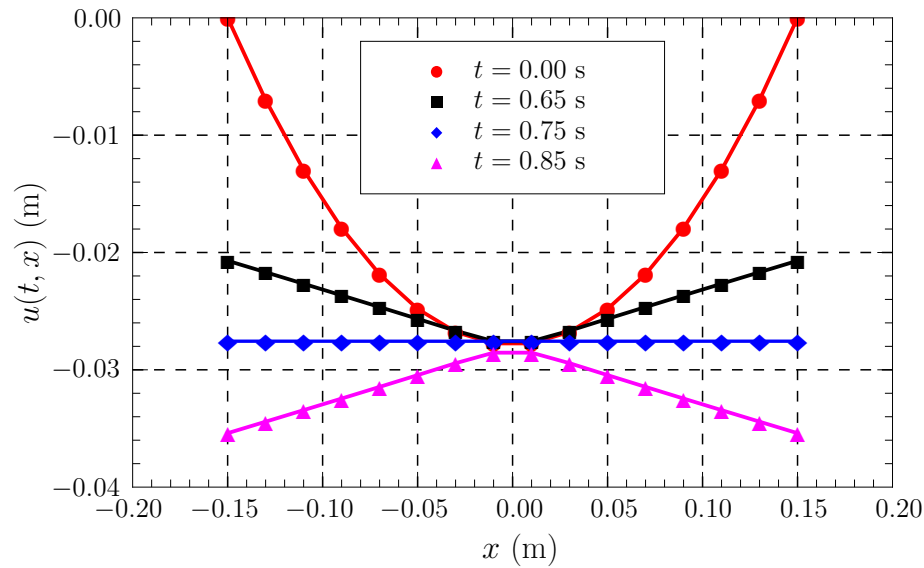}
\caption{Numerical results of the falling catenary. Until the signal arrives the middle point levitates. Used parameters are the same with that of the \fref{fig:catenaryfall1}.}
\label{fig:catenaryfall2}
\end{figure}

The motion can be numerically simulated and the results are given at \fref{fig:catenaryfall1} and \fref{fig:catenaryfall2}. As we expected, the middle point of the catenary levitates until the signal arrives. Note that in this case the falling speed is not constant. Each point falls with constant acceleration during the time for the levitation of the middle point. In this case, the signal direction and the material's falling direction is not parallel, but still, until this signal arrives, the point maintains its current state of motion. We can say that the falling spring is to a longitudinal wave what the falling catenary is to a transverse wave \cite{ref:longitudinal, ref:transverse}. For a homogeneous transverse wave, the oscillating motion has nothing to do with the wave speed. That is why the material can be accelerating in the falling catenary case in contrast to the falling spring case.

From the two cases, we can conclude that the elastic wave phenomenon is the key to the understanding of the elapsed time for the levitation. That is, the effect of the application or the negation of a force transfers along with the wave and it takes a finite time. In addition to that, depending on the situation, signal propagation speed is relevant to the movement speed of the composite part of the bulk.

\section{Demonstration}
\begin{figure}
\centering
\includegraphics[width=1.0\textwidth]{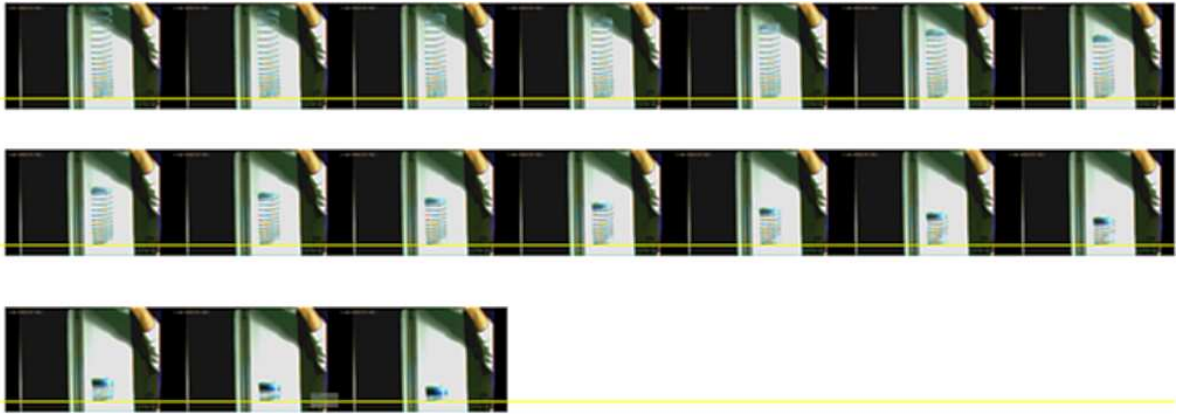}
\caption{The falling spring demonstration. Frame interval is 1/125 sec. See the bottom point. The top point falls with constant speed.}
\label{fig:demo}
\end{figure}

We here demonstrate the falling spring only. It will be intriguing to see if the falling catenary is tested. We have recorded the motion of the falling spring by using a high speed digital camera. For a high school or undergraduate level lab, a normal digital camera will be enough \cite{ref:camera}. \Fref{fig:demo} shows consecutive frames. Each frame interval is 1/125 sec. Relevant measured values are given at \tref{tab:table1}. The speed of the top point $v_{top}$ was getting by counting the pixel of the each frame. We also calculated the value of the initially stretched length $l$ by using \eref{eq:springl}. This is same order of magnitude with the measured value. The order of signal speed for the falling spring is given by $v_s \simeq l \sqrt{3k/m}$ to the observer on the ground. Here the factor $\sqrt{3}$ came from the fact that our experimental set up is closer to the tightly wound spring case. In that case the characteristic time is order of $\sqrt{m/(3k)}$, not $\sqrt{m/k}$ \cite{ref:spring2}. By blindly inserting the measured $m, k$ and $l_0$ values to \eref{eq:springsol}, we cannot get the graph like \fref{fig:springfall}. Instead it will predict crossed lines which mean inversion of spatial order of spring which is impossible. This is because for our parameters the loosely wound spring approximation is too simple and we should consider the thickness of the spring and the pretension. But still this approximation helps to capture the qualitative behaviour and our result shows this behaviour too. We can see clearly that the bottom point levitates. And the top point falls with constant speed which is smaller than the signal speed. We think this tabletop demonstration is a reasonable method for introducing the force transferring phenomenon in an evident way.

\begin{table}[tbp]
\centering
\caption{The falling spring experimental data. $v_{top}$ is a falling speed of the top point. $v_s$ is a signal speed given by $ l \sqrt{3k/m}$.}
\label{tab:table1}
\begin{tabular}{l l l @{} l @{} l l @{.} l @{} l }
\br
\multicolumn{1}{c}{Parameter} &
\multicolumn{1}{c}{Unit} &
\multicolumn{3}{c}{Measured} &
\multicolumn{3}{c}{Calculated} \\
\mr
$m$ & \,\, kg & 3&.10 & $\times 10^{-2}$  \\
$k$ & \,\, N/m & 9&.8 & $\times 10^{-1}$  \\
$l_0$ & \,\, m & 3&.6 & $\times 10^{-2}$ \\
$l$ & \,\, m & 3&.0 & $\times 10^{-1}$ & 1&9 & $\times10^{-1}$ \\
$v_{top}$ & \,\, m/s & 2&.15 & \\
$v_s$ & \,\, m/s &  &   & & 2&9 & \\
\br
\end{tabular}
\end{table}

\section{Summary and Discussion}
Force requires time to be transferred. This fact is not well tasted in problem solving if we approximate the bulk as a particle. So at undergraduate courses where the particle model is heavily used, this fact is barely used. In this paper we contrast this fact by the numerical simulation and the demo experiment. The falling spring and the falling catenary both show the levitation behaviour of the bottom points. This is due to the very fact that the force requires time to propagate. And the negation of the force also requires time to propagate through the system.

A bulk is in the static equilibrium means that at each point of the bulk the composite forces balance each other to give zero net force. When this equilibrium is broken at one point, this unbalance propagates through the bulk. Naturally, this propagation takes the form of the wave and the mechanical wave carries the information. Until this information is arrived, the other points do not know the application or negation of the force, so it keeps its equilibrium state. This is the physical argument for the levitation of the falling spring and falling catenary.

Mechanical waves take mainly two forms, longitudinal and transverse. For the longitudinal wave through the solid bulk, the signal speed usually becomes the limiting speed. In other words, the composite material cannot move faster than the signal speed or there must be a spatial inversion or acausality. Falling spring shows this fact by the constant falling speed. For a homogeneous transverse wave, the wave speed has nothing to do with the oscillating motion. So, for the falling catenary, the composite part can show the accelerating motion. But still the signal propagates with the finite speed.

These rich physics are barely caught when we replace the bulk as a particle. Even though the particle approximation is inevitable in teaching the entry level physics, this should be regarded as a replaced concept to be. Once acquainted with the particle concept, it is needed to let the student see the bulk as a continuum with a finite size. And it is recommended to mention that there is a more involved physics even in the simple set up. This will help the students from getting wrong thinking habits and will enlarge their view points. The evident features of the falling spring and the falling catenary make them as concrete examples for the force transferring phenomena. Naturally, the force transfer is related with the subject of wave. So it can be emphasized further that the wave is a generic behaviour of the nature. But the discussion about this point is another subject and we leave it for the future work.

\ack
The authors thank two students, Minseok Sim and Ungsik Nahm for their help for the experiment. This work was supported by the Korea Science Academy of KAIST and Korea Foundation for the Advancement of Science \& Creativity (KOFAC). 

\end{document}